\documentclass[aps,pra,amsfonts,amsmath,twocolumn,superscriptaddress]{revtex4}
\usepackage{amsmath}
\usepackage{amssymb}
\usepackage{multirow}
\usepackage{graphicx}
\usepackage{color}
\usepackage[normalem]{ulem}
\usepackage{float}
\newcommand{\bse}{\begin{subequations}}
\newcommand{\ese}{\end{subequations}}
\newcommand{\be}{\begin{equation}}
\newcommand{\ee}{\end{equation}}
\newcommand{\bea}{\begin{eqnarray}}
\newcommand{\eea}{\end{eqnarray}}
\newcommand{\kp}{\kappa}
\newcommand{\lam}{\lambda}
\newcommand{\al}{\alpha}

\begin{document}


\title{Material Dependence of the Wire-Particle Casimir Interaction}



\author{E. Noruzifar}
\affiliation{Department of Physics and Astronomy,
  University of California, Riverside, California 92521, USA}

\author{P. Rodriguez-Lopez}
\affiliation{{Departamento de F\'isica Aplicada I and GISC, Facultad de Ciencias F\'isicas, Universidad Complutense, 28040 Madrid, Spain}}
\affiliation{{Departamento de Matematicas and GISC, Universidad Carlos
    III de Madrid, Avenida de la Universidad 30, 28911 Legan\'es,
    Spain}}

\author{T. Emig}
\affiliation{Laboratoire de Physique Th\'eorique et Mod\`eles
 Statistiques, CNRS UMR 8626, Universit\'e Paris-Sud, 91405 Orsay,
 France}

\author{R. Zandi}
\affiliation{Department of Physics and Astronomy,
  University of California, Riverside, California 92521, USA}


\date{\today}


\begin{abstract}
  We study the Casimir interaction between a metallic cylindrical wire
  and a metallic spherical particle by employing the scattering
  formalism.  At large separations, we derive the asymptotic form of the
  interaction. In addition, we find the
  interaction between a metallic wire and an isotropic atom, both in
  the non-retarded and retarded limits.
{We identify the conditions under which 
the asymptotic Casimir interaction does not depend on the material 
properties of the metallic wire and the particle.}
Moreover, we compute the exact Casimir interaction between the
particle and the wire numerically. We show that there is a complete
agreement between the numerics and the asymptotic energies at large
separations. For short separations, our numerical results show good
agreement with the proximity force approximation.
\end{abstract}

\maketitle


%

\section{Introduction}
\label{sec:intro}

Casimir forces contribute significantly to the effective interaction
of micro- and nanometer sized structures \cite{CasimirPhysics}.  For
identical objects or mirror symmetric configurations, this type of
interaction is attractive \cite{Kenneth:2006uq} and can cause stiction
in micro-motors and other similar structures \cite{serry98}. 
More generally, if the permittivities of the objects are higher or
lower than those of the surrounding medium, any equilibrium position of
the objects is unstable due to the Casimir interactions \cite{Rahi:2010kx}.
Therefore, a good quantitative understanding of such forces is a key
parameter in the design and manufacturing of micro-mechanical devices.

It is important to study the Casimir forces for different shapes as
they strongly depend on the geometry and material properties
\cite{kardar99,umar,CasimirPhysics}.  Technically, investigating the
interplay between the shape and material effects is quite involved.
The scattering formalism provides a powerful tool to calculate the
Casimir interaction between objects of general shape and material
properties \cite{universal}.  There is much recent research activity
based on the scattering formalism, e.g., for edges and tips
\cite{Maghrebi:2011kx, Graham:2010uq}, anisotropic particles
\cite{Emig:2009fk}, wires and plates \cite{ours, ours1,emig06,
  rahi08}, and spheres and plates \cite{roya10}.

An important geometry which has not yet been investigated in detail,
consists of a wire and a particle (atomic or macroscopic). In the
plane-particle geometry this force is known as Casimir-Polder (CP)
interaction \cite{casimir}. Our study of the wire-particle case is
motivated by theories \cite{Arnecke:2008vn} and experiments
\cite{Denschlag:1998ys,Bawin:2001zr} on the two-dimensional quantum
scattering of neutral atoms or molecules at wires or nanotubes.  In an
early work, the interaction between a filament and an isotropic atom
has been studied for perfectly and non-perfectly conducting metals
\cite{barash89}.  Later, Eberlein and Zietal studied the interaction
between a neutral atom and a perfect metal cylinder, using
perturbation theory \cite{eberlein1, eberlein2}.  Recently, the
Casimir energy for a polarizable micro-particle and an ideal metal
cylindrical shell has been computed using the Green's function
technique \cite{sc-mostep}.
The focus of previous studies were mainly on the interaction between a
 metal wire and a perfect metal particle or an
atom. Therefore, the influence of material properties of the spherical
particles on the energy remains to be studied in detail.

In this work, we study the Casimir interaction between a metallic
spherical particle and cylindrical metallic wire where the latter is
described by the Drude, plasma or perfect metal model.  Using the
scattering formalism, we derive a general expression for the Casimir
interaction between the particle and cylinder.  From this general
expression we determine the behavior of the Casimir energy in various
limiting cases (separation regimes) analytically, and numerically over
a wide range of separations.  Interestingly, we find ranges of
distances in which the Casimir interaction does not depend on the
material properties of the metallic wire.  In contrast, we find that
the interaction depends in general on the material properties of
the metallic particle at all separations. An exception is the
  plasma sphere with a plasma wavelength that is much smaller than the
  size of the sphere for which the Casimir interaction is universal at
  asymptotically large distances.
At short separations, we compute the exact Casimir interaction
numerically and compare it both with the asymptotic results and the
prediction of proximity force approximation (PFA). In both limits we
obtain good agreement.

The structure of the rest of the paper is as follows: In
Sec.~\ref{sec:formalism}, we review the scattering approach and derive
the elements that are needed for computing the interaction between a
wire and a particle.  In Sec.~\ref{sec:largeseparation}, the
large--distance asymptotic interaction between a metallic wire and a
particle (metal sphere and isotropic atom) is derived for the perfect
metal, plasma and Drude models.  In Sec.~\ref{sec:numerics}, the exact
Casimir interaction is computed numerically and compared with the
asymptotics expansions.  Section
\ref{sec:pfa} is dedicated to the interaction  at short separations
where the PFA is expected to become reliable.

\section{Method}
\label{sec:formalism}

We consider a cylindrical wire separated by a distance $d$ from a
spherical particle.  We use the scattering formalism to calculate the
Casimir energy between the cylinder and sphere \cite{universal}.  In
general, the Casimir {free} energy between two objects { at the
  temperature $T$} is given by
\be
\label{eq:energyT}
{\mathcal E} = k_{B} T {{\sum_{n=0}^{\infty}}^{\prime}} \ln\,{\det}
\left[{\bf 1} - \mathbb{N}(\kp_n) \right]\,,
\ee
where $\bf 1$ is the identity matrix, 
$\kp_n$ is the Matsubara wave number $\kp_n = 2\pi n k_{B}T/\hbar c$ and 
the matrix $\mathbb N$ factorizes into the scattering amplitudes (T-matrices) 
as well as translation matrices which describe the coupling between the 
multipoles on distinct objects. 
The primed sum denotes that the contribution of $n=0$ has to be weighted by 
a factor of $1/2$. 

At zero temperature, the primed sum in Eq.~\eqref{eq:energyT} 
is replaced by an integral along the imaginary frequency axis,
\be
\label{eq:energy}
{\mathcal E} = \frac{\hbar c}{2\pi} \int_0^{\infty} d\kp \ln\det \left( {\bf 1} - \mathbb{N}\right)\,,
\ee
with $\kp$ the Wick-rotated frequency. 
The elements of the matrix $\mathbb N$ for electric ($E$) and magnetic  ($M$) polarizations ($\alpha,~\beta,~\gamma=E,~M$)
and cylindrical wave functions $m$ and $m'$ are 
\begin{multline}
\label{eq:nmatrix}
{\mathbb N}_{k_z m, k_z' m'}^{\alpha\beta}=\sum_{\gamma=E,M}\sum_{{m''}=-\infty}^{\infty}T_{{\rm s},k_z m ,k_z' m''}^{\alpha\gamma}\, \\
\times\sum_{n=-\infty}^{\infty}
{\mathcal U}^{\rm sc}_{k_z' m''n}\, T_{{\rm c},k_z'n}^{\gamma\beta}\, {\mathcal U}^{\rm cs}_{{k_z'}nm'}\,,
\end{multline}
where $k_z$ is the wave number along the 
$z$-axis$,T_{\rm c}$  and $T_{\rm s}$ are the T-matrices of the cylinder and sphere in 
cylindrical basis, respectively. 
The translation matrix $\mathcal U^{\rm sc}$ 
relates regular cylindrical vector waves to outgoing ones. 

The translation matrices do not couple different polarizations 
and for both $E$ and $M$ polarizations, their matrix elements 
are given by
\begin{align}
  \label{eq:trans}
 {{\mathcal U}^{\rm sc}_{n n'}}&=(-1)^{n'}\, K_{n-n'} (p\,d) \,,\nonumber\\
{\mathcal U}^{\rm cs}_{n n'}& = (-1)^{n-n'} {\mathcal U}^{\rm sc}_{n n'}\,,
 \end{align}
where $p = \sqrt{\kappa^2+k_z^2}$ and $K_n(x)$ is the modified Bessel function of the second kind. 
Note that the $\mathbb N$-matrix elements in 
Eq.~\eqref{eq:nmatrix} are written in cylindrical basis 
to avoid the complicated form of 
the translation matrices in the spherical basis \cite{universal}. 

The T-matrix of the sphere in cylindrical basis is derived in Appendix \ref{app:s-tmat-in-cyl} and is given by
\bea
\label{eq:textTsctoTss}
T_{{\rm s},k_z m,k_z'm}^{{\alpha\gamma}} = 
\frac{1}{2\pi \kp L}\sum_{ \ell ={\rm max}(1,|m|)}^{\infty}\sum_{{\beta}} (1-2\delta_{{\alpha,\beta}})& \nonumber\\
\times D^{\dagger}_{k_z m {\alpha}, \ell m {\beta}}  \,T_{{\rm s}, \ell m}^{{\beta}}\,
D_{\ell m{\beta},k_z' m {\gamma}}\,,&
\eea
where $L$ is the length of the cylinder, $\ell$ is the 
quantum number of the spherical electromagnetic waves, 
${\beta}=E,~M$ is the electromagnetic polarization  and $D$ is the conversion 
matrix from the cylindrical to spherical basis. 
The elements of the conversion matrix are given in Appendix \ref{conversion} . 
Note that in Eq.~\eqref{eq:textTsctoTss}, $T_{\rm s,\ell m}^{{\beta}}$ is the 
T-matrix of the sphere in the spherical basis. 

To obtain the Casimir energy from Eq.~\eqref{eq:energy}, 
we plug Eqs.~\eqref{eq:trans} and \eqref{eq:textTsctoTss} into \eqref{eq:nmatrix} and  
use the identity 
$\det({\bf 1}-{\bf AB})= \det({\bf 1}-{\bf BA})$. 
The $\mathbb N$-matrix for the energy 
between the sphere and the cylinder is rewritten as
\begin{multline}
\label{eq:N'}
{\mathbb{N}}_{\ell m, \ell' m'}^{{\alpha \beta}}= \frac{1}{4\pi^2\kp}\sum_{{\gamma,\gamma'}} T_{{\rm s},\ell m}^{{\alpha}} 
\int_{-\infty}^{\infty} dk_z\, {D}_{\ell m {\alpha},k_z m {\gamma}} \,
 \,\\
 \times{\bf\tilde T}_{mm'}^{{\gamma\gamma'}} {D}^{\dagger}_{k_z m' {\gamma'},\ell' m'{\beta}} (1-2\delta_{{\beta,\gamma'}})
\,,
\end{multline}
with
\be
{\bf\tilde T}_{mm'}^{{\gamma\gamma'}}=\sum_{n=-\infty}^{\infty} 
{\mathcal U}_{k_z mn {\gamma}}^{\rm cs} \, T^{{\gamma \gamma'}}_{{\rm c},k_z n} 
\, {\mathcal U}_{k_z nm' {\gamma'}}^{\rm sc}\,.
\ee
In this work, 
to study the impact of the material properties of the metallic objects
 on the Casimir interaction, 
we employ the 
plasma, Drude and 
perfect metal dielectric properties 
with the constant magnetic permeability $\mu= 1$. 
The Drude model dielectric response is given by 
\be
\label{di-fun}
\epsilon (i c \kappa,\lambda_{\rm p},\lambda_{\sigma}) = 
1+ \frac{(2\pi)^2}{(\lambda_{\rm p}\kappa)^2+\lambda_{\sigma}\kappa/2}\,,
\ee
where $\lambda_{\rm p}$ is the plasma wave length and 
$\lambda_{\sigma} = 2\pi c/\sigma$ is the length scale associated with the 
conductivity $\sigma$. 
Equation~(\ref{di-fun}), 
reproduces the plasma model dielectric function  for $\lambda_{\sigma}=0$.
Note that the material properties of the 
sphere and the cylinder enter to the calculations 
through the T-matrices, see Appendix \ref{tmatrix}. 
\section{Large-separation regime: asymptotic Casimir energy}
\label{sec:largeseparation}
In this section, we study the large separation asymptotic behavior of 
the Casimir interaction between a particle and a wire. 
We consider a spherical particle with radius $R_{\rm s}$ and 
a cylindrical wire with radius $R_{\rm c}$. 
In order to find the large--distance ($d \gg R_{\rm c}, R_{\rm s}$)
asymptotic form of the Casimir interaction, 
one has to find the behavior of the T-matrices in {the low-frequency} limit.

\subsection{Asymptotic Behavior of T-matrices}
\label{subsec:Tasym}
\subsubsection{T-matrix of a wire}
In this part, we obtain the
asymptotic form of the 
 T-matrix elements of a wire at large separations. 
Using the dielectric function given in Eq.~\eqref{di-fun}, we find the T-matrix element of the 
wire for $E$ polarization and $n=0$ at small frequencies ($\kappa\ll 1,~ k_z/\kappa ~{\rm fixed}$) is %
\be
\label{tcasym}
T^{EE}_0 \approx -\frac{p^2 }{C(\kp)- p^2 
 \ln (p R_{\rm c})}\,,
\ee
where $p=\sqrt{\kappa^2+k_z^2}$. The parameter $C(\kp)$ depends on the 
dielectric properties of the wire.  For a perfect metal wire $C(\kappa) = 0$ 
and for a plasma wire with the plasma wavelength $\lambda_{\rm p}$, 
$C(\kp)\approx \lambda_{\rm p}^2 \kp^2/(2\pi^2 R_{\rm c}^2)$ 
if the plasmon oscillations cannot build up transverse to the wire axis as 
the diameter is too small, i.e., $R_{\rm c} \ll \lambda_{\rm p}$. 
In the opposite limit we approximately reproduce the T-matrix of a perfect metal wire,  
i.e. $C(\kp)\approx0$. For a Drude wire with the conductivity $\sigma$ 
and the characteristic length $\lambda_\sigma$, $C(\kp)=\lambda_\sigma \kp/(4\pi^2R_{\rm c}^2)$ 
if $\kp \ll \lambda_{\sigma}/\lambda_{\rm p}^2, 1/\lambda_{\sigma}$.  
The first condition ($\kp \ll \lambda_{\sigma}/\lambda_{\rm p}^2$) means that the Drude behavior dominates over the plasma 
behavior, equivalent to the fact that in the dominator of Eq.~\eqref{di-fun}, 
the first term is much smaller than the second one. The second condition ($\kappa\ll 1/\lambda_{\sigma}$)
ensures that the Drude dielectric function is much larger than one, i.e., the metallic 
behavior is dominant \cite{ours, ours1}.\\
At large separations,  $T^{EE}_0$ elements dominate over
the other T-matrix elements since $T^{EM}_0=T^{ME}_0=0$, 
$T^{MM}_0 \sim \kp^2$,  and 
for {$n\ne0$} partial waves $ T_n\sim \kp^{2|n|}$. Note that 
 for Drude cylinders $T^{EE}_0\sim \kp$ while for  
plasma and perfect metal cylinders $T^{EE}_0\sim 1$.
\subsubsection{T-matrix of a particle}
The T-matrix elements of a spherical particle have a 
different scaling  compared to the ones for the cylindrical wire. 
For the plasma and perfect metal spheres 
$T_{{\rm s},\ell m}\sim\kp^{2\ell+1}$. 
For Drude spheres the T-matrix elements for 
the electric $E$ and magnetic $M$ polarizations 
scale differently: $T^{M}_{{\rm s}, \ell m}\sim \kp^{2(\ell+1)}$ 
and $T^{E}_{{\rm s}, \ell m}\sim\kp^{2\ell+1}$. 
Therefore, the asymptotic behavior of the T-matrix 
at large separations is 
dominated by the $\ell=1$ elements. 
The asymptotic form of the T-matrix elements for the magnetic polarization $M$ 
and $\ell =1$ depends on the material properties of the sphere. 
While for a perfect metal sphere, we have
\be
\label{pftm}
T_{{\rm s},1m}^M \approx -\kp^3 R_{\rm s}^3/3\,,
\ee
for a plasma sphere with the plasma wavelength $\lambda'_{\rm p}$, we obtain
\be
\label{pltm}
T_{{\rm s},1m}^M \approx 
-\frac{1}{3}\frac{I_{5\over2}(2\pi R_{\rm s}/\lambda'_{\rm p})}{I_{1\over2}(2\pi R_{\rm s}/\lambda'_{\rm p})}
\kp^3 R_{\rm s}^3\,.
\ee
Note that in the limit of perfect conductivity $\lambda_{\rm p}'\to 0$, we reproduce 
Eq.~\eqref{pftm}. 

For a Drude sphere with the conductivity $\sigma'$ and 
the characteristic length $\lambda'_\sigma$,  
\be
\label{drtm}
T_{{\rm s},1m}^M \approx -\frac{8\pi^2}{45}\frac{R_{\rm s}}{\lambda'_{\sigma}}\kp^4 R_{\rm s}^4\,.
\ee
However, the asymptotic form of the T-matrix elements for $E$ polarization and $\ell=1$, 
up to the leading order, does not 
depend on the material properties, and is the same for the 
perfect metal, plasma and Drude models,
\be
\label{tseasym}
T_{{\rm s},1m}^E \approx \frac{2}{3} \kp^3 R_{\rm s}^3\,.
\ee
Note that for the Drude sphere $T_{{\rm s},1m}^M \sim \kp^4$, 
and thus we add the sub-leading term to 
the expansion of $T_{{\rm s},1m}^E$ in Eq.~\eqref{tseasym} and obtain

\be
\label{tseDasym}
T_{{\rm s},1m}^E \approx \frac{2}{3} \kp^3 R_{\rm s}^3 -\frac{1}{4\pi^2}\frac{\lambda_\sigma'}{R_{\rm s}}\kp^4 R_{\rm s}^4\,,
\ee
where the sub-leading term contains the material properties of the Drude wire.
\subsubsection{T-matrix of an Atom}
The above approach can also be used to calculate the Casimir energy between an atom and a wire. 
To this end, we consider a neutral two-level atom in the ground state, 
with the transition frequency  $\omega_{10}$ \cite{casimir}. We assume 
that the distance from the atom to the wire $d$ is much larger 
than the radius of the wire $R_{\rm c}$, i.e. $d/R_{\rm c}\gg 1$.
 Moreover, we assume that the atom is isotropic and 
does not have magnetic polarizability. 
In the isotropic-dipole approximation, the only nonzero element of the 
T-matrix reads 
\be
\label{eq:atom-tmatrix}
T^{E}_{{\rm atom},1m}\approx \frac{2}{3} \alpha^E(\kappa) \,{\kappa^3}\,,
\ee
where $\alpha^E$, the electric polarizability, is given by
\begin{equation}
\label{eq:electric-polarizability}
 \alpha^{E}(\kappa) = \frac{\alpha_0}{1+\kappa^2 d_{10}^2}\,,
\end{equation}
with $d_{10}=c/\omega_{10}$,  $\alpha_0=f_{10}e^2/(m\omega_{10}^2)$, $e$ the electron charge, $m$ the mass and $f_{10}$  the 
oscillator strength of the $0\rightarrow 1$ transition.

\subsection{Asymptotic Energy Expression}
\label{asym:energy}
In this subsection, using the asymptotic T-matrix expressions and
Eq.~\eqref{eq:N'}, we derive the Casimir energy at large
separations.  Considering ${\ln{\rm det}\equiv {\rm
      Tr}\ln}$, we expand the integrand in Eq.~(\ref{eq:energy})
{in powers of $\mathbb{N}$} for
$\kp\ll 1$ and $k_z/\kp$ fixed and find
\be
\label{eq:energykzexp}
{\mathcal E} \approx -\frac{\hbar c}{2\pi} \int_0^{\infty} d\kp 
\,{\rm Tr} {{\mathbb{N}}}\,.
\ee
As discussed above, in the limit $d\gg R_{\rm s}, R_{\rm c}$ only the $\ell=1$ terms contributes to
the sphere T-matrix, $T_{{\rm s},\ell m}$.
Therefore, only the
partial wave numbers $m=0,\pm 1$ {have} to be taken into
account to obtain the matrix ${\mathbb N}$.  Using Eq.~(\ref{eq:N'}), we find
\begin{multline}
\label{thetrace}
{\rm Tr}\,{\mathbb N}  \approx \sum_{\gamma}\sum_{m=-1}^{1}\tilde{\mathbb{N}}_{1 m, 1 m}^{{\gamma\gamma}} \\
= \frac{1}{4\pi^2\kp}\sum_{{\gamma}}\sum_{m=-1}^{1} T_{{\rm s},1 m}^{{\gamma}} 
\int_{-\infty}^{\infty} dk_z\,  {\tilde D}_{1 m {\gamma},k_z m E}\,
{\bf\tilde T}_{m {m}}^{EE}\, \\
 \times {\tilde D}^{\rm T}_{k_z m E, 1 m {\gamma}} (1-2\delta_{{\gamma},E})\,,
\end{multline}
with $\tilde{\bf  D}$ the modified conversion matrix with real elements. 
The modified conversion matrix is related to the original one by 
${\bf D}_{\ell m,k_z m}=(-1)^{\ell-m}(-i)^{\ell+m-1} {\tilde{\bf D}}_{\ell m,k_z m}$ 
, see Appendix \ref{conversion}. Furthermore, the matrix ${\bf\tilde T}_{mm}^{EE}$ in Eq.~(\ref{thetrace}) is equal to 
\be
\label{tildet}
{\bf\tilde T}_{mm}^{EE}\approx K_{m}^2(p d) T^{EE}_{{\rm c},k_z 0}\,.
\ee
Note that in Eq.~(\ref{tildet}), $T^{EE}_{{\rm c},k_z\pm 1}$ are neglected as they scale with higher 
powers of $\kp$. 

Inserting Eq.(\ref{tildet}) into Eq.~(\ref{thetrace}) and performing the sums, we find
\begin{multline}
\label{thetrace2}
-{\rm Tr}[\tilde{\mathbb N}] \approx 
\frac{1}{4\pi^2 \kp} \int_{-\infty}^{\infty} dk_z\,T^{EE}_{{\rm c},k_z 0}\\
\times \left[
K_0^2(p d) ({\tilde D}_{10E,k_z0M}^2 T^{M}_{{\rm s},1} -{\tilde D}_{10M,k_z0M}^2 T^{E}_{{\rm s},1}) \right.\\
\left. +2 K_1^2(p d) ({\tilde D}_{11E,k_z1M}^2 T^{M}_{{\rm s},1} -{\tilde D}_{11M,k_z1M}^2 T^{E}_{{\rm s},1})\right]\,.
\end{multline}
The modified conversion matrix elements in Eq.~\eqref{thetrace2} are (see Appendix \ref{conversion} for all details ),
\begin{eqnarray}
\label{eq:conv10}
{\tilde D}_{10E,k_z0M}&=&0, ~~~~~~ {\tilde D}_{10M,k_z0M}=\sqrt{6\pi}(1+k_z^2/k^2)^{1/2}, \notag\\
{\tilde D}_{11E,k_z1M}&=&\sqrt{3\pi},~~ {\tilde D}_{11M,k_z1M}=\sqrt{3\pi}k_z/k. 
\end{eqnarray}

Inserting Eq.~\eqref{eq:conv10} into Eq.~(\ref{thetrace2}) and using the leading term in {the} 
$T^{E}_{{\rm s},1 m}$ expansion, we find 
the asymptotic energy between a perfect metal, plasma and Drude {cylinder} and a spherical particle or an atom. 
{ Using Eq.~\eqref{eq:energykzexp} and the T-matrix expansions up to $\kappa^3$, 
see Eqs.~\eqref{pftm}-\eqref{tseDasym}, we obtain the general expression for the asymptotic energy}
\begin{multline}
\label{eq:asym-int}
\frac{\mathcal E}{\hbar c} \approx
\frac{1}{\pi^2 }
\int_{0}^{\infty} d\kp
\int_{0}^{\infty} dk_z\,
T^{EE}_{{\rm c},0}\\
\times \chi\left(
p^2 K_0^2(p d) + (2\Lambda_0 \kappa^2 + k_z^2) K_1^2(p d)\right) \,,
\end{multline}
with $\chi = R_{\rm s}^3$ for the spherical particle and $\chi = \alpha^E(\kappa)$ for the isotropic atom. 
Moreover, $\Lambda_0$ is proportional to $T^{M}_{{\rm s},1}$, see Eq.~\eqref{thetrace2}, with 
$\Lambda_0 = 1/4$ for the perfect metal particle, 
$\Lambda_0=
{I_{5\over2}(2\pi R_{\rm s}/\lambda'_{\rm p})}/({4I_{1\over2}(2\pi R_{\rm s}/\lambda'_{\rm p})})$
for the plasma particle and $\Lambda_0 = 0$ for the atom and the Drude particle.
{ The 
latter is due to the fact that $T^M_{{\rm s},1}$ is set to zero in Eq.~\eqref{eq:asym-int}. 
$T^M_{{\rm s},1}$ for the atom is indeed zero and 
for the Drude particle scales with $\kappa^4$, see Eq.~\eqref{drtm}. 
Therefore, for the Drude particle 
the asymptotic energy given by Eq.~\eqref{eq:asym-int} 
needs a correction because of the $\kp^4$ terms in Eqs.~\eqref{drtm} and \eqref{tseDasym}, which is}
\begin{multline}
\frac{\delta{\mathcal E}_{\rm DS}}{\hbar c} =
\frac{1}{\pi^2}
\int_{0}^{\infty} \kp\, d\kp
\int_{0}^{\infty} dk_z\,
T^{EE}_{{\rm c},0} \\
\times \chi\left( \frac{3\lambda_\sigma'}{8\pi^2} p^2 K_0^2 (p d) +
\left[ \frac{3\lambda_\sigma'}{8\pi^2} k_z^2 - \frac{4\pi^2 R_s^2}{15\lambda_\sigma'}\kp^2\right] K_1^2(p d)
\right)\,.
\end{multline}
Inserting 
Eq.~(\ref{tcasym}) into Eq.~(\ref{eq:asym-int}) and 
using the polar coordinates, $\kp \to \rho\cos(\theta)/d$ and $k_z\to \rho \sin(\theta)/d$, 
we find
\begin{multline}
\label{eq:asym-int-tm0-pol}
\frac{\mathcal E}{\hbar c} \approx 
 -\frac{1}{\pi^2 d^4\ln (2d/R_{\rm c}) }
\int_{0}^{\infty} d \rho \rho^3 
\int_{0}^{\pi/2} \frac{d\theta}{1+C(\rho,\theta)} \\
 \times \chi\left( K_0^2(\rho) + (\sin^2(\theta)+2\Lambda_0\cos^2(\theta)) K_1^2(\rho) \right) 
\,,
\end{multline}
where  for the 
perfect metal cylinder $C(\rho,\theta)=0$, 
for the plasma cylinder $C(\rho,\theta)\approx \xi\cos^2(\theta)$ with 
$\xi = {{\lambda_{\rm p}^2} /(2\pi^2 R_{\rm c}^2}\ln (2d/R_{\rm c}))$, 
and for the Drude cylinder $C(\rho,\theta)=\xi' \cos(\theta)/\rho$ 
with $\xi'=\lambda_\sigma d/(4\pi^2 R_{\rm c}^2\ln(2d/R_{\rm c}))$. 

The correction  
 $\delta{\mathcal E}_{\rm DS}$ for the Drude particle in polar coordinates $(\rho,\theta)$ reads
\begin{multline}
\label{eq:corr}
\frac{\delta{\mathcal E}_{\rm DS}}{\hbar c} =
\frac{3 R_{\rm s}^3\,\lambda_\sigma'}{8\pi^4 d^5 \ln(2d/R_{\rm c})}
\int_{0}^{\infty} \rho^4\, d\rho
\int_{0}^{\pi/2} d\theta\,
\frac{\cos(\theta)}{1+C(\rho,\theta)} \\
\times \left(K_0^2 (\rho) +
\left[ \sin^2(\theta) - \frac{32\pi^4 R_s^2}{45\lambda_\sigma'^2}\cos^2(\theta)\right] K_1^2(\rho)
\right)\,.
\end{multline}

Now we use Eqs.~\eqref{eq:asym-int-tm0-pol} and 
\eqref{eq:corr} for different material properties and 
calculate the Casimir interaction for various limiting cases. 

{\subsection{ The Casimir interaction between a wire and a spherical particle}
\label{subsec:wire-particle asymptotics}

Below we present the large separation asymptotic energies between a 
metallic spherical particle and a metallic wire. 
}

\subsubsection{The perfect metal wire}
For a perfect metal wire and a perfect metal, plasma or Drude particle, 
the energy integral in Eq.~(\ref{eq:asym-int-tm0-pol}) results into 
\be
\label{eq:asym-cyl-pf-sph-npf}
\frac{\mathcal E}{\hbar c} \approx -\frac{\Lambda_0+1}{3\pi}\frac{R_\text{s}^3}{d^4\ln(2d/R_{\rm c})}\,.
\ee
It is important to note that Eq.~\eqref{eq:asym-cyl-pf-sph-npf} depends on the 
material properties of the spherical particle through the quantity
$\Lambda_0$.  For a perfect metal wire and a plasma particle, 
in the limiting case of small plasma wavelengths, ${\lambda'}_{\rm p}\ll R_{\rm s}$, 
we reproduce the perfect conductivity form with ${{\Lambda}_0} \approx 1/4$. 
In the limit of large plasma wavelengths, ${\lambda'}_{\rm p}\gg R_{\rm s}$, 
we obtain $\Lambda_0 \approx {\pi^2 R_{\rm s}^2}/{(15{\lambda'}_{\rm p}^2)}$. 
Since $\Lambda_0 \ll 1$, the plasma wavelength of the spherical particle does not 
have a significant contribution to the asymptotic energy.

For the Drude particle, using Eq.~\eqref{eq:corr}, the correction to 
the asymptotic energy given by 
Eq.~\eqref{eq:asym-cyl-pf-sph-npf} reads 
\be
\label{eq:pf-corr}
\frac{\delta {\mathcal E}_{\rm DS}}{\hbar c} =- \frac{R_{\rm s}^3}{32\,d^5\,\ln(2d/R_{\rm c})}
\left(
\frac{\pi^2 R_{\rm s}^2}{2\, \lambda_\sigma'}-\frac{63\, \lambda_\sigma' }{64\, \pi^2 }
\right)\,,
\ee
and {scales} with $d^{-5}$. 
Depending on the terms in the 
{brackets}, this correction can have a significant contribution to the asymptotic energy.
In the limit of high conductivity, $\lambda_\sigma' \ll R_{\rm s}$,
the first term in the 
{brackets} dominates over the second one. For this specific case, if $\lambda_\sigma' d \ll R_{\rm s}^2$  
the correction becomes even larger than the asymptotic energy itself, i.e. $\delta {\mathcal E}_{\rm DS} \gg  {\mathcal E}$. 

Note that in Eq.~\eqref{eq:pf-corr}, the second term in the brackets 
dominates only at low conductivity limit $\lambda_\sigma' \gg R_{\rm s}$ in which the spherical particle is considered to be a very poor conductor. In general, the second term does not have a 
noticeable contribution to the asymptotic energy for good conductors such as copper and gold.

\subsubsection{The plasma wire}
For a plasma wire and a plasma or a perfect metal particle, 
the polar and radial integrals in Eq.~(\ref{eq:asym-int-tm0-pol}) 
can easily be performed, 
\be
 \frac{\mathcal E}{\hbar c} \approx -\frac{R_\text{s}^3}{\pi d^4 \ln(2d/R_{\rm c})}
f(\xi)\,,
\ee
with
\be
f(\xi)=\frac{1}{3 \xi}(2\Lambda_0-1)\left(1-(1+\xi)^{-{1\over2}}\right)
+\frac{1}{2}(1+\xi)^{-{1\over2}}\,,
\ee
{where $\xi$ is given below Eq.~\eqref{eq:asym-int-tm0-pol}, $\xi = \lambda_{\rm p}^2/(2\pi^2 R_{\rm c}^2 \ln(2d/R_{\rm c}))$.}
In the small plasma wavelength limit, $\lambda_{\rm p}/R_{\rm c}\ll \sqrt{\ln(2d/R_{\rm c})}$, 
we reproduce the perfect metal wire results given by Eqs.~(\ref{eq:asym-cyl-pf-sph-npf}) and ~\eqref{eq:pf-corr}.

In the opposite limit $\lambda_{\rm p}/R_{\rm c}\gg \sqrt{\ln(2d/R_{\rm c})}$, 
the asymptotic energy reads
\begin{multline}
\label{eq:asym-cyl-pl}
\frac{\mathcal E}{\hbar c} \approx 
-\frac{R_{\rm c}R_\text{s}^3}{\lambda_{\rm p} d^4}
 \left(\frac{1}{\sqrt{2\ln(2d/R_{\rm c})}}
+\frac{4\pi\Lambda_0}{3}\frac{R_{\rm c}}{\lambda_{\rm p}}\right)\,, 
\end{multline}
For a perfect metal spherical particle, $\Lambda_0=1/4$, and the second term 
can be neglected as we are at the large plasma wavelength regime. 

For a plasma spherical particle with plasma wavelength $\lambda_{\rm p}'$, 
if $\lambda_{\rm p}'\ll R_{\rm s}$, $\Lambda_0 \approx 1/4$ and 
the energy is the same as in the case of 
a perfect particle and a plasma wire. In the opposite limit 
 $\lambda_{\rm p}'\gg R_{\rm s}$, we have $\Lambda_0 \approx {\pi^2 R_{\rm s}^2}/{(15{\lambda'}_{\rm p}^2) \ll 1}$ and the plasma wavelength of the particle does not 
have a significant 
{effect on} the asymptotic energy, and the asymptotic energy is mainly dominated by the 
material properties of the plasma wire.

For a Drude particle, the asymptotic energy is given by 
Eq.~\eqref{eq:asym-cyl-pl} together with a correction obtained by Eq.~\eqref{eq:corr}. 
The correction reads
\begin{equation}
\label{eq:pl-corr}
\frac{\delta {\mathcal E}_{\rm DS}}{\hbar c}=-
\frac{3}{64}\frac{R_{\rm s}^3}{d^5} \left(\frac{R_{\rm c}}{\lambda_{\rm p}}\right)^2
\left(
 \frac{9}{4} (\frac{5}{8}-\ln2)\lambda_\sigma' 
 +\pi^4\frac{ R_{\rm s}^2}{\lambda_{\sigma}'}
\right)\,.
\end{equation}
Since for good conductors
$\lambda_\sigma' \ll R_{\rm s}$, similar to the case of a perfect metal wire, 
the second term dominates over the first one.
\subsubsection{The Drude wire}
For a Drude wire and a perfect metal or plasma particle, 
in the limit $\xi'\ll 1$ or $d^2/R_{\rm c}^2 \ll d/\lambda_\sigma$, 
using Eqs.~\eqref{eq:asym-int-tm0-pol} and \eqref{eq:corr}, 
 we reproduce the perfect metal wire results, see Eqs.~(\ref{eq:asym-cyl-pf-sph-npf}) and 
\eqref{eq:pf-corr}.

In the opposite limit, $\xi' \gg 1$ or equivalently $d^2/R_{\rm c}^2 \gg d/\lambda_\sigma$ 
and $d\gg\lambda_{\rm p}^2/\lambda_\sigma$
the integrations in Eq.~(\ref{eq:asym-int-tm0-pol}) result into 
\bea
\label{eq:asym-cyl-dr-sph}
\frac{\mathcal E}{\hbar c} \approx 
-\frac{9\pi^2}{64}\frac{R_\text{s}^3 R_{\rm c}^2}{\lambda_{\sigma} d^5}
\left(5\Lambda_0+4\ln({\lambda_\sigma d}/{R_{\rm c}^2})\right)
\,.
\eea
For a perfect metal particle, $\Lambda_0=1/4$, and the energy 
given by Eq.~\eqref{eq:asym-cyl-dr-sph} is always attractive since 
$\lambda_\sigma d \gg R_{\rm c}^2$.

For the plasma particle,
in the limit of small plasma wavelengtha $\lambda_{\rm p}'\ll R_{\rm s}$, 
$\Lambda_0\approx 1$ and the particle behaves like a perfect metal. In the opposite 
limit, $\lambda_{\rm p}'\gg R_{\rm s}$, as previously seen,
we find $\Lambda_0 \approx {\pi^2 R_{\rm s}^2}/{(15{\lambda'}_{\rm p}^2)}$. 
In this case the second term in Eq.~\eqref{eq:asym-cyl-dr-sph} dominates, which means that the 
material properties of the plamsa particle does not have a significant 
{effect on} 
the asymptotic Casimir energy.

Under the same condition 
{ $\xi'\gg 1$}, using Eq.~\eqref{eq:corr}, 
the correction to Eq.~\eqref{eq:asym-cyl-dr-sph} reads
\be
\label{eq:corr-dr}
\frac{\delta{\mathcal E}_{\rm DS}}{\hbar c}=
-\frac{R_{\rm c}^2\, R_{\rm s}^3 \ln(2d/R_{\rm c})}{\lambda_\sigma d^6}
\left(
\frac{32\pi^3 R_{\rm s}^2}{25\lambda_\sigma'}-\frac{7\lambda_\sigma'}{5\pi}
\right)\,.
\ee

As discussed above, for metallic 
particles, the second term 
in Eq.~\eqref{eq:corr-dr} is much smaller than the first one. 
{
\subsection{The Casimir energy between a wire and an Atom}
\label{subsec:wire-atom}
We calculate the Casimir energy between a wire and an atom in both retarded and non-retarded limits. To find the asymptotic energies at large separations, we use Eq.~\eqref{eq:asym-int-tm0-pol} with $\Lambda_0=0$: 

\subsubsection{The retarded limit}

In the retarded limit, $d\gg d_{10}$, we find the Casimir energy between a
perfect metal wire and an atom as
\begin{equation}
 \label{eq:atom-perf-wire}
 \frac{\mathcal E}{\hbar c} \approx -\frac{1}{3\pi}
 \frac{\alpha_0}{d^4 \ln(2d/R_\text{c})}\,.
\end{equation}
%

Equation~\eqref{eq:atom-perf-wire} is in complete agreement with the 
results in Refs.~\cite{eberlein1,eberlein2,sc-mostep,barash89}. 
It is important to note that even though the asymptotic energies for an atoms given in Eq.~\eqref{eq:atom-perf-wire} 
and for a perfect metal particle given in Eq.~\eqref{eq:asym-cyl-pf-sph-npf}
have the same scaling behavior, the
numerical coefficients do not match; $1/(3\pi)$ for the atom and $5/(12\pi)$ for the spherical particle.
This discrepancy is due to the lack of magnetic polarizability in the isotropic atoms. 

For the plasma wire and an atom in the limit $\xi \gg 1$ or equivalently 
$\lambda_\text{p}/R_\text{c} \gg \sqrt{\ln(2d/R_\text{c})}$ and $d\gg d_{10}$, 
the asymptotic energy reads
\begin{equation}
 \label{eq:atom-pl-wire}
 \frac{\mathcal E}{\hbar c} \approx - \frac{R_\text{c}}{\lambda_\text{p}}\frac{\alpha_0}{d^4\sqrt{2\ln(2d/R_\text{c})}}\,,
\end{equation}
which is in agreement with the atom--plasma wire result in Ref.~\cite{barash89}. 
In the limit $\lambda_\text{p}/R_\text{c} \ll \sqrt{\ln(2d/R_\text{c})}$ and $d\gg d_{10}$, 
we reproduce the perfect metal wire-atom interactin energy given in Eq.~\eqref{eq:atom-perf-wire}. 

For the Drude wire and an atom, in the region of intermediate distances $\xi' \gg 1$  or equivalently $d/R_{\rm c} \gg R_{\rm c}/\lambda_\sigma$, 
in the retarded limit $d\gg d_{10}$, 
the asymptotic energy reads
\be
\label{eq:atom-dr-wire}
\frac{\mathcal E}{\hbar c} \approx 
-\frac{9\pi^2}{16}\frac{\alpha_0 R_{\rm c}^2}{\lambda_{\sigma} d^5}
\ln({\lambda_\sigma d}/{R_{\rm c}^2})
\,.
\ee
Equation~\eqref{eq:atom-dr-wire} is in agreement with Ref.~\cite{barash89}. 
In the opposite limit, $\xi' \ll 1$ or $d/R_{\rm c} \ll R_{\rm c}/\lambda_\sigma$ and $d\gg d_{10}$, 
once again we reproduce the perfect metal wire-atom asymptotic interaction energy, see Eq.~(\ref{eq:atom-perf-wire}).

\subsubsection{The non-retarded limit}

In the non-retarded limit, $d\ll d_{10}$,  
using Eq.~\eqref{eq:asym-int-tm0-pol} with $\Lambda_0=0$, 
we perform the angular integral for the perfect metal, plasma 
and Drude wires. 

For the perfect metal wire, the radial integral can easily be obtained.
Expanding the result of the integral for $d\ll d_{10}$ yields
\begin{equation}
\label{eq:atom-pf-nonr}
\frac{\mathcal E}{\hbar} \approx-\frac{\pi}{16} \frac{\alpha_0\,\omega_{10}}{d^3 \,\ln(2d/R_\text{c})}\,.
\end{equation}

For the plasma wire, the radial integral over $\rho$ in Eq.~\eqref{eq:asym-int-tm0-pol} 
cannot readily be performed in the non-retarded limit. Therefore, we expand the integrand 
for $d/d_{10} \ll 1$. 
In the limit $d \ll R_\text{c} d_{10} / \lambda_\text{p}$ 
the expansion of the integrand does not depend on the material 
properties up to the leading order. Therefore, performing the integration over $\rho$ 
results into the perfect metal wire-atom interaction energy
given in Eq.~\eqref{eq:atom-pf-nonr}.

Similar to the plasma wire, the radial integration in Eq.~\eqref{eq:asym-int-tm0-pol}  is not easily calculable for the 
Drude wire. Analogously, we expand the integrand for $d/d_{10} \ll 1$ and then perform the 
integral over $\rho$. In the limit $d \ll R_\text{c} \sqrt{ d_{10} / \lambda_\sigma }$, 
we find Eq.~\eqref{eq:atom-pf-nonr} for a perfect metal wire and an atom. 
This is due to the fact that the material properties of the Drude wires does not play any role ate intermediate separations, 
see Refs.~\cite{ours1,ours}.
}
\subsection{Universality}

In previous sections, we have derived the asymptotic energies
between a metallic wire and a metallic spherical particle for different dielectric
properties, described by the Drude, plasma or perfect metal models.
We have found that in all cases, the Casimir energy depends on the
material properties of the spherical particle. This is due to the fact that 
the material property of a sphere has 
a significant contribution to its T-matrix \cite{roya10}.

In contrast, for parallel metallic wires and a wire--plate geometry, 
{at intermediate distances,}
the asymptotic energy does not depend on the material properties of {the} objects and is universal \cite{ours,ours1}.

\begin{figure}
\includegraphics[width=0.5\textwidth]{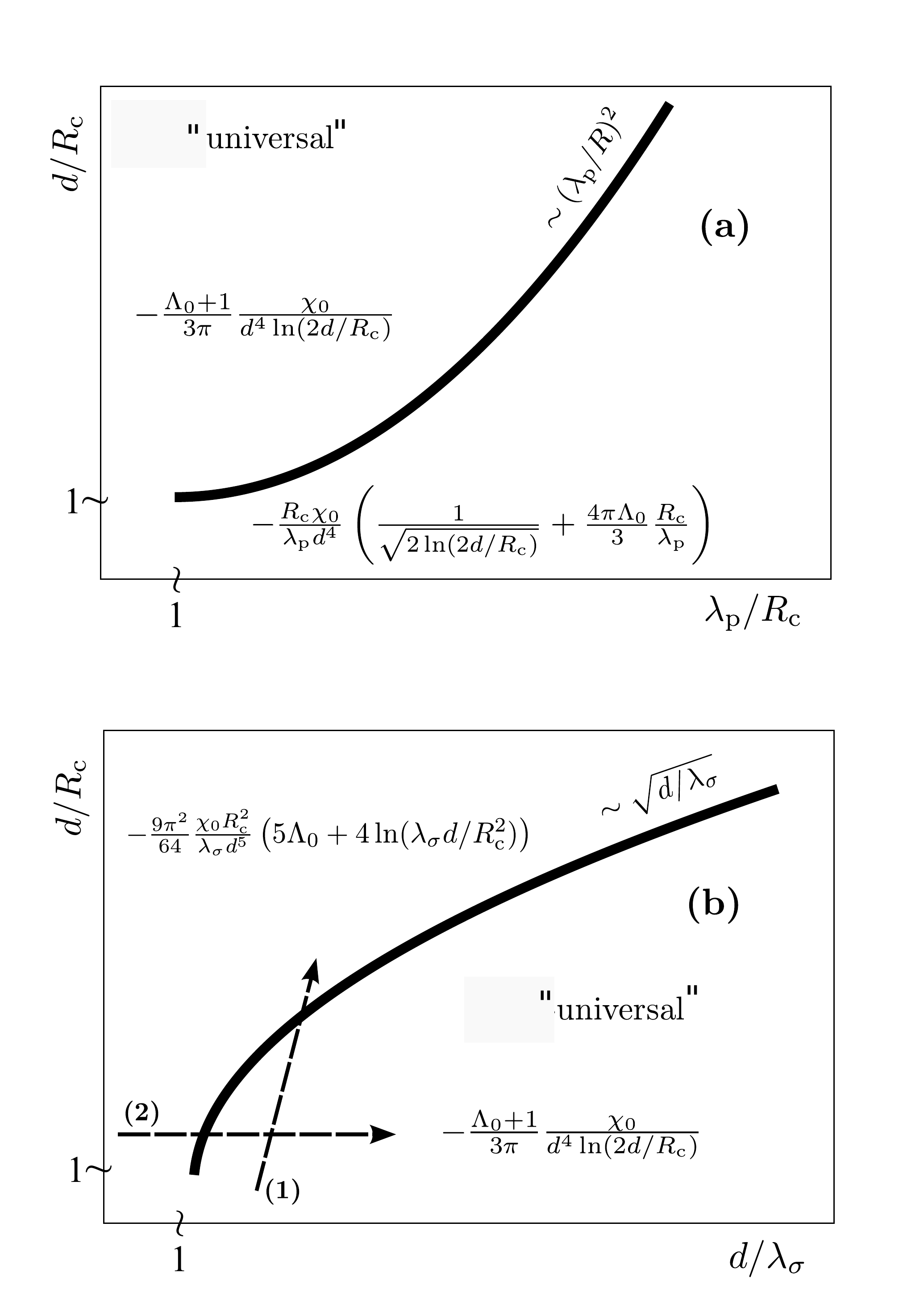}
\caption{\label{fig:semi-universal} 
  Interaction between a wire and a particle and between a wire and an atom 
  in the retarded limit. {The formulas describe} the
  rescaled interaction energies, ${\cal E}/(\hbar c)$. 
  The parameter $\chi_0=\alpha_0$ for the atom and $\chi_0=R_\text{s}^3$ for 
  the particle. (a) 
  Interaction {between} a plasma wire {and} a metallic spherical particle. The asymptotic
  results corresponds to the regimes sufficiently far from the 
  curve $\ln(d/R_{\rm c})\sim (\lambda_{\rm p}/R_{\rm c})^2$ and for $d/R_{\rm c}$, $\lambda_{\rm p}/R_{\rm c} \gg
  1$. (b) Interaction {between} a Drude wire {and} a metallic particle. 
  The separating curve is given, up to logarithmic corrections, by
  $d/R_{\rm c} \sim \sqrt{d/\lambda_\sigma}$. The energy expressions hold for
  $d/R_{\rm c}$, $d/\lambda_\sigma \gg 1$ and $d \gg
  \lambda_{\rm p}^2/\lambda_\sigma$. Different regimes can be reached depending on the relative {size} of length scales: Arrow (1)
  corresponds to an increasing distance $d$ which ultimately leads to
  a strictly {\it non-universal} interaction. Arrow (2) indicates an
  overall increase of the geometry (i.e., $d/R_{\rm c}$ fixed) with constant
  conductivity leading to a `` {\it universal} '' interaction.
}
\end{figure}

Although in a wire--sphere system 
the Casimir interaction depends on the 
material properties of the particle at all separations, 
the signatures of the universal behavior of the metallic wire 
is still traceable at asymptotic separations. 
For a plasma wire at intermediate distances, 
$d/R_{\rm c}\ll \exp(\lambda_{\rm p}^2/R_{\rm c}^2)$, 
the Casimir interaction depends both on 
the material properties of the wire and the particle. 
For larger separations, $d/R_{\rm c}\gg \exp(\lambda_{\rm p}^2/R_{\rm c}^2)$, 
the interaction is independent of the 
material properties of the plasma wire, 
while {it} still depends on the material properties of the particle, {see Fig.\ref{fig:semi-universal}a}.


For a Drude wire and a metallic particle, at larger separations, $d^2/R_{\rm c}^2\gg d/\lambda_{\sigma}$, 
the interaction depends both  
on the material properties of the Drude wire and that of the spherical particle, 
see Fig.\ref{fig:semi-universal}b.
At intermediate distances $d^2/R_{\rm c}^2\ll d/\lambda_{\sigma}$, while material properties of the sphere has a significant role in the Casimir energy,
it does not depend on the material properties of the wire.  Note that the casimir interaction is also independent of the material properties for two parallel Drude wires \cite{ours,ours1}.

\section{Intermediate-separation regime: numerical calculations }
\label{sec:numerics}

We use Eq.~\eqref{eq:energy}, to numerically calculate the Casimir energy. 
The numerical algorithm consists of three major parts: 
(i) constructing the matrix ${{\mathbb N}}$ from Eq.~(\ref{eq:N'}), 
(ii) computing the determinant of ${\bf 1}-{{\mathbb N}}$ for 
specific imaginary frequencies $\kp$ and (iii) 
integrating over $\kp$. 
The matrix ${\mathbb N}$ 
consists of blocks which are associated 
with the quantum numbers $\ell$ and $\ell'$. 
We truncate $\ell$ and $\ell'$ at a finite partial wave number 
$\ell_{\rm max}$ such that the result for the energy changes 
by less than a factor of $1.0001$ upon increasing $\ell_{\rm max}$ 
by $1$. Since $\ell,\ell'\leq\ell_{\rm max}$, the N-matrix has $\ell_{\rm max}^2$ blocks {$N_{\ell\ell'}$}.
%
%
The block ${{\bf N}}_{\ell\ell'}$ consists of {the elements} ${{\bf N}}_{\ell m,\ell'm'}$ 
which form $2\times2$ blocks,  
with $m=-\ell,\dots, \ell$ and $m'=-\ell',\dots, \ell'$,
\be
\nonumber
{ {\bf N}}_{\ell m, \ell' m'}=\begin{pmatrix}
 {{\mathbb N}}^{MM}_{\ell m, \ell' m'} & {{\mathbb N}}^{ME}_{\ell m, \ell' m'}\\
 {{\mathbb N}}^{EM}_{\ell m, \ell' m'} & {{\mathbb N}}^{EE}_{\ell m, \ell' m'}
\end{pmatrix}\,.
\ee
\begin{figure}
\includegraphics[width=0.45\textwidth]{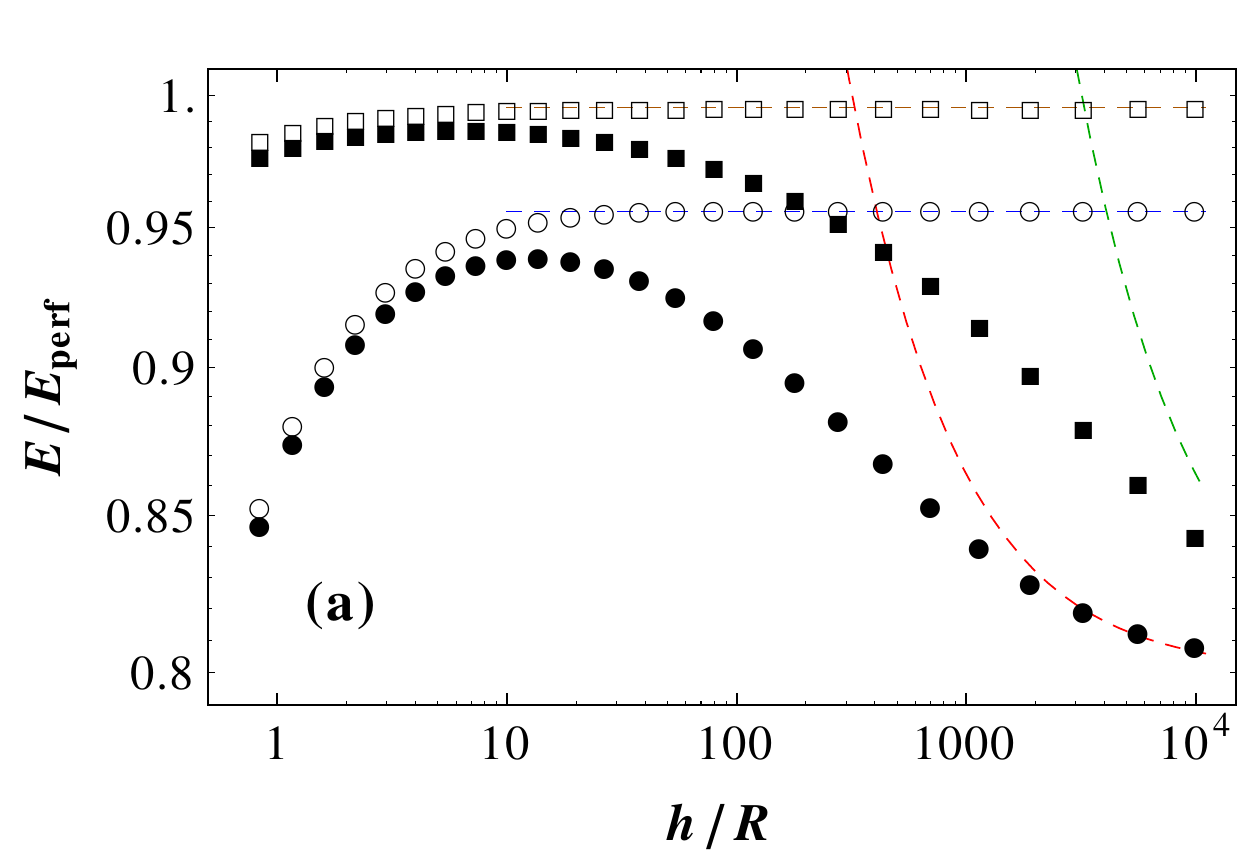}
\includegraphics[width=0.45\textwidth]{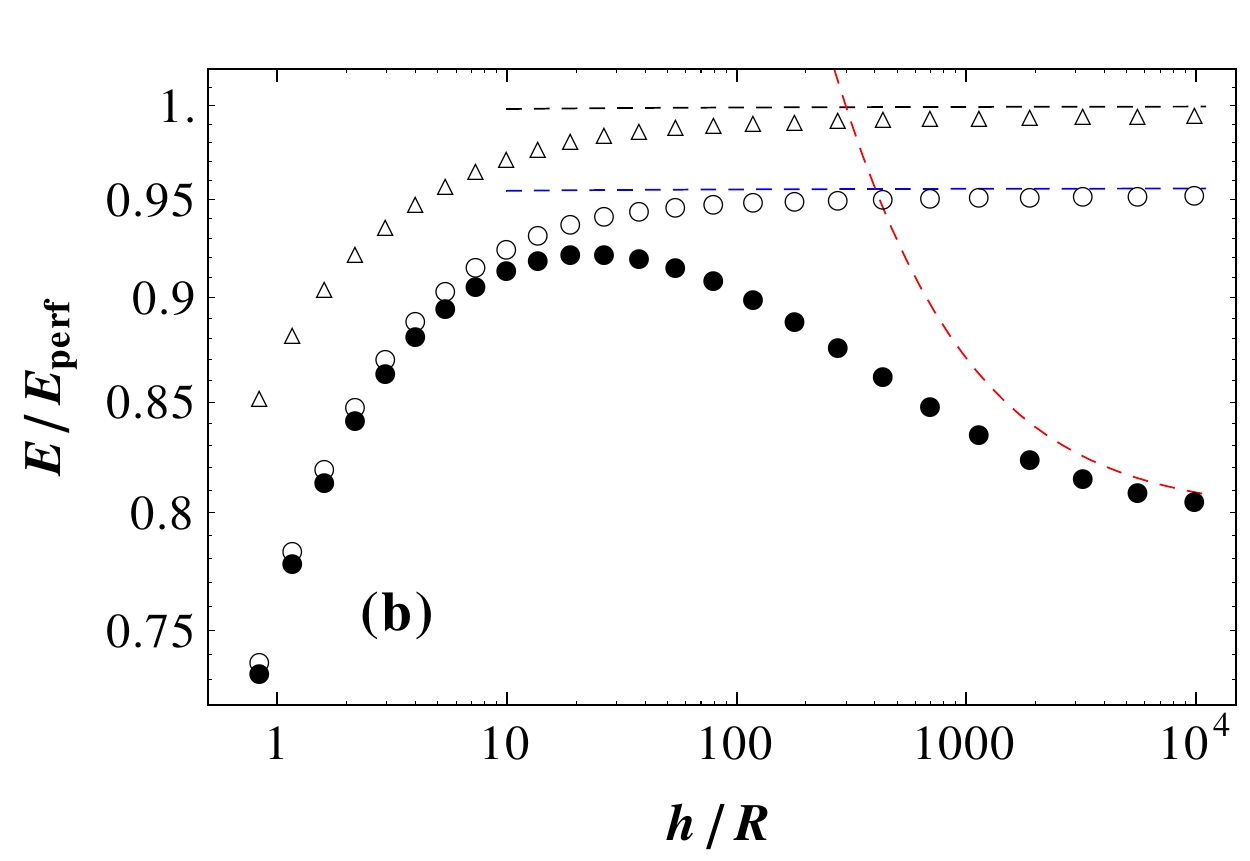}
\includegraphics[width=0.45\textwidth]{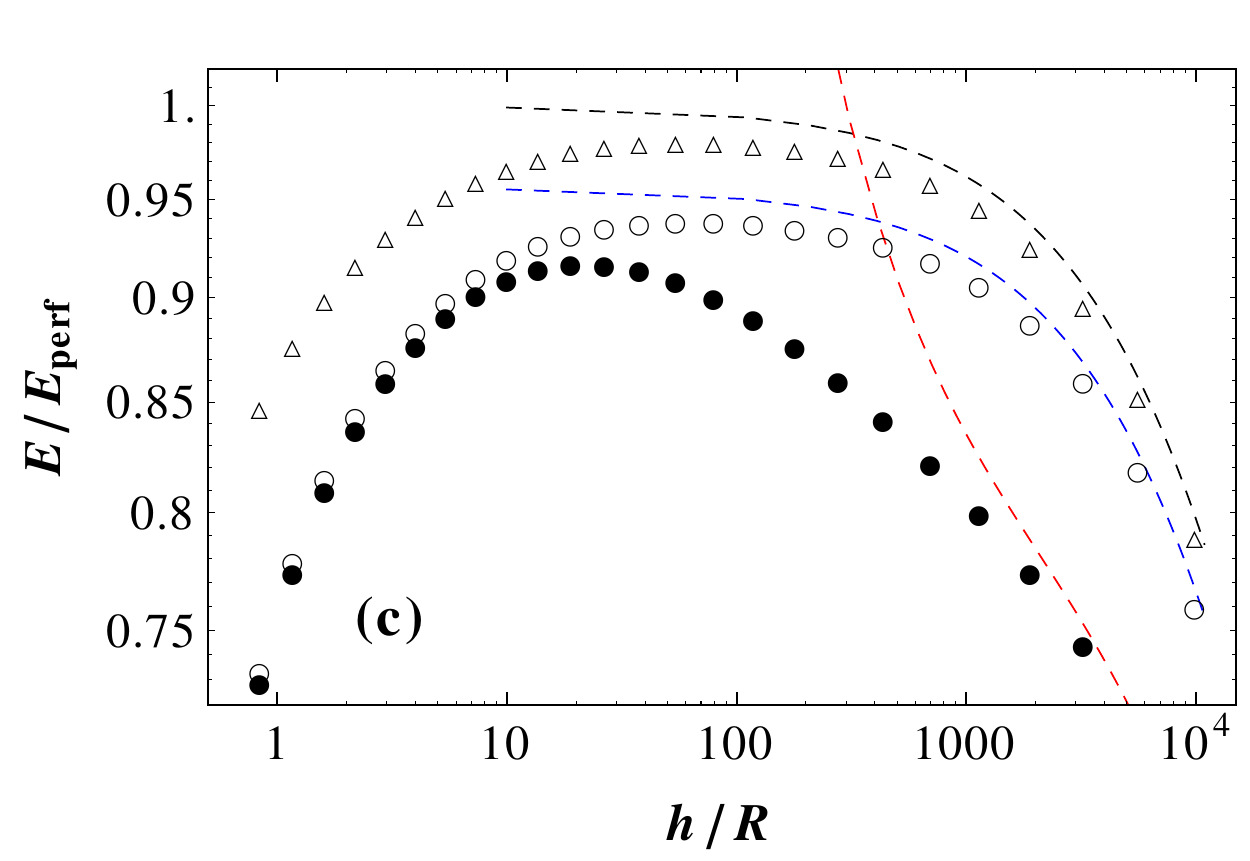}
\caption{ \label{fig:numerics} 
Ratio of the numerically computed energy $E$ for realistic metals 
to the perfect metal energy $E_{\rm perf}$ versus $h/R$ for a wire and a particle with $R_\text{c}=R_\text{s}=R$.
Each figure shows the numerics related to a wire with 
specific material properties: (a) for a perfect metal wire; 
(b) a plasma wire with $\lambda_{\rm p}/R=0.5$ and 
(c) a Drude wire with the same value for $\lambda_{\rm p}/R$ and $\lambda_{\sigma}=\lambda_{\rm p}/27.4$. 
The ratio of energies for particles with different material properties are distinguished 
by different symbols: Open triangles for 
perfect metal particle and open circles and open squares for plasma 
particles with $\lambda_{\rm p}/R=0.5,~ 0.05$, respectively. 
Filled circles and filled squares denote Drude particles with $\lambda_{\rm p}/R=0.5,~0.05$, respectively,
 and $\lambda_{\sigma}=\lambda_{\rm p}/27.4$.}
\end{figure}
Consequently the size of the block ${{\bf N}}_{\ell\ell'}$ is $(4\ell+2)\times (4\ell'+2)$, 
implying that the off-diagonal blocks ($\ell\neq\ell'$) are not square matrices. 
Furthermore, ${\mathbf N}_{\ell\ell'}$ blocks are not { diagonal} since
symmetry along the axis parallel to the wire's axis is broken by the particle.

To construct the matrix ${{\mathbb N}}$, $4\ell^2_{\rm max}(\ell_{\rm max}+2)^2$ integrals over $k_z$ 
have to be evaluated for each $\kappa$. 
This makes the numerical computations for closer separations quite expensive. For example at separation $d/R=2.6$, 
the energy converges with $\ell_{\rm max}=12$, corresponding to a matrix of size $336$ with
$112896$ $k_z$-integrals for just a single {value} of $\kappa$. 

Figure \ref{fig:numerics} illustrates our numerical results for a
metallic {cylinder} 
and a metallic spherical particle both with the radius $R$.  The plots
show the Casimir energy normalized to the energies for the perfect
metal wire-particle configuration as a function of the surface-to-surface
distance $h=d-2R$.  For the numerical calculations, we used
$\lambda_{\rm p}/R = 0.05$ and $0.5$ with $\lambda_{\rm
  p}/\lambda_{\sigma}=27.4$, corresponding to the parameters for 
gold with $\lambda_{\rm p}=137$ nm and $\lambda_{\sigma}\approx
5$ nm \cite{gold}.  Figure \ref{fig:numerics} shows the dependence of
the Casimir energy {on} the material
properties of the wire and particle.  Figure \ref{fig:numerics}(a)
depicts the interaction energy between the perfect metal wire with the metallic particle
for plasma ({open}
symbols) and Drude (filled symbols) models with $\lambda_{\rm p}/R=0.05$
(squares) and $0.5$ (circles) and $\lambda_{\rm
  p}/\lambda_{\sigma}=27.4$.  The dashed lines show the asymptotic
energies given by Eqs.~\eqref{eq:asym-cyl-pf-sph-npf} and
\eqref{eq:pf-corr}.  As shown in the figure, there is a very good agreement
between the asymptotics and the numerical results.  The numerical
results confirm that at very large separations, only the material
properties of the particle contribute to the Casimir energy.
{We note that in the range of distances considered here and for
  the parameters of gold, one has to consider the
  energy given in Eq.~\eqref{eq:asym-cyl-pf-sph-npf} in addition to the correction term presented in
  Eq.~\eqref{eq:pf-corr}.}

Figure \ref{fig:numerics}(b) shows the interaction between a plasma
wire and a perfect metal
({open} triangles), plasma
({open} circles) and
Drude particles (filled circles) with the plasma wavelength
$\lambda_{\rm p}/R = 0.5$ and $\lambda_{\rm p}/\lambda_{\sigma}=27.4$.
The dashed lines show the asymptotic energies given by
Eqs.~\eqref{eq:asym-cyl-pl} and \eqref{eq:pl-corr}. This figure also
shows the good agreement between the asymptotics derived in the
previous section and our numerical results.  {Again, here it is
  important to include the correction given by
  Eq.~\eqref{eq:pl-corr}.}  Note that for the perfect metal particle
and the plasma cylinder ({open} 
triangles) the energy ratio approaches 1 at large separations.  This
is due to the fact that in this regime the material properties of the
plasma wire do not contribute to the Casimir interaction, see
Fig.~\ref{fig:semi-universal}(a).

Figure \ref{fig:numerics}(c) shows the interaction between a Drude wire 
and a perfect metal ({open} triangles), a plasma ({open} circles) and 
a Drude spherical particles (filled circles) with $\lambda_{\rm p}/R=0.5$ and $\lambda_{\rm p}/\lambda_{\sigma}=27.4$. The dashed lines are obtained by computing the integrals 
in Eqs.~\eqref{eq:asym-int-tm0-pol} and \eqref{eq:corr}. 
Since for $10^2\lesssim d\lesssim10^4$ 
we have $10^{-2}\lesssim \xi'\lesssim10^{-1}$, corresponding to 
the crossover regime, 
the asymptotic energ{ies} {of Eqs.~\eqref{eq:asym-cyl-dr-sph} and \eqref{eq:corr-dr}} {are} not applicable, see Fig.~\ref{fig:semi-universal}(b).

\section{Short-separation regime: Proximity Force Approximation}
\label{sec:pfa}

\begin{figure}
\hspace{-.7cm}
\includegraphics[width=0.45\textwidth]{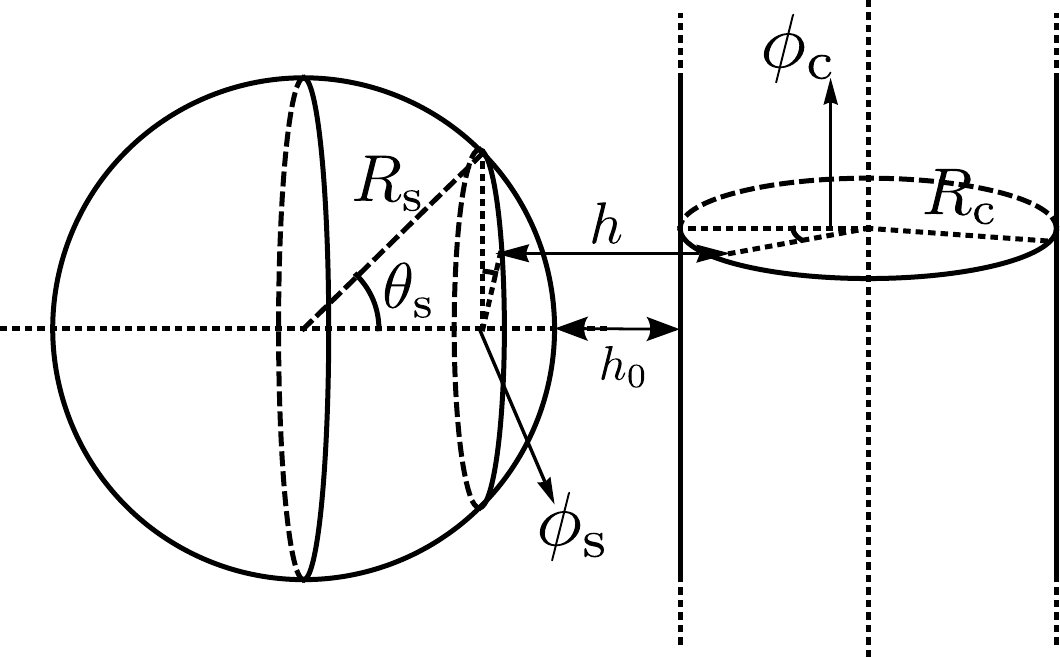} 
\caption{{The geometry of a wire of radius $R_c$ and a sphere of radius $R_s$ at a distance $h_0$ (center-to-center distance $d=h_0+R_c+R_s$).
}}
\label{pfa-figure}
\end{figure}

In this section, using the Proximity Force Approximation 
(PFA) \cite{derj}, 
we calculate the Casimir interaction at short separations 
$h\ll R_{\rm c},R_{\rm s}$. This method gives the interaction as  
{an integral} of the energies between parallel {surface segments, }
\be
\label{pfa-general}
{\mathcal E}_{\rm PFA}=\int dA \,E_{\rm plate}(h)\,,
\ee
where $E_{\rm plate}$ is the Casimir energy per unit area between 
two parallel plates and $h$ 
is the surface-to-surface distance. 
Figure~\ref{pfa-figure} illustrates the distance between two surface elements. 
According to Fig.~\ref{pfa-figure} the distance $h$ is  
\be
h = h_0 + R_{\rm s}(1-\cos(\theta_{\rm s}))+R_{\rm c}(1-\cos(\phi_{\rm c}))\,,
\ee
with $h_0=d-R_{\rm c}-R_{\rm s}$ the distance of the closest approach between the cylinder and the sphere. 
One can write $\phi_{\rm c}$ in terms of $\theta_{\rm s}$ and $\phi_{\rm s}$,
\be
\sin(\phi_{\rm c})=\frac{R_{\rm s}}{ R_{\rm c}} \sin(\theta_{\rm s})\sin(\phi_{\rm s})\,.
\ee
At short separations, the surface elements of the sphere and cylinder in which $\theta_{\rm s}\ll 1$ and $\phi_{\rm c}\ll 1$ respectively,  contribute most {to} the interaction. Therefore, the distance $h$ 
can be approximated by
%
%
\be
\label{stos}
h(\theta_{\rm s},\phi_{\rm s}) \approx h_0 + \frac{R_{\rm s}\theta_{\rm s}^2}{2} 
( 1+\frac{R_{\rm s}}{R_{\rm c}}\sin^2(\phi_{\rm s}))\,.
\ee
Inserting Eq.~(\ref{stos}) into Eq.~(\ref{pfa-general}) and 
performing a simple change of variable, we obtain the PFA energy  
%
%
\be
\label{eq:pfa-integral}
{\mathcal E}_{\rm PFA} = \frac{2\pi\,{ R_{\rm s}}}{\sqrt{1+\frac{R_{\rm s}}{R_{\rm c}}}}
\int_{h_0}^{\infty}  d {H}\, 
E_{\rm plate}({H})\,.
\ee
For the case of perfect metal {surfaces} with $E_{\rm plate}(H)/A=-\pi^2\hbar c/(720 H^3)$, we find
\be
{\mathcal E}_{\rm PFA} =-\frac{\pi^3\,{R_{\rm s}}\,\hbar c}{720\sqrt{1+\frac{R_{\rm s}}{R_{\rm c}}}\,{h_0^2}}\,.
\ee
\begin{figure}
\includegraphics[width=0.4\textwidth]{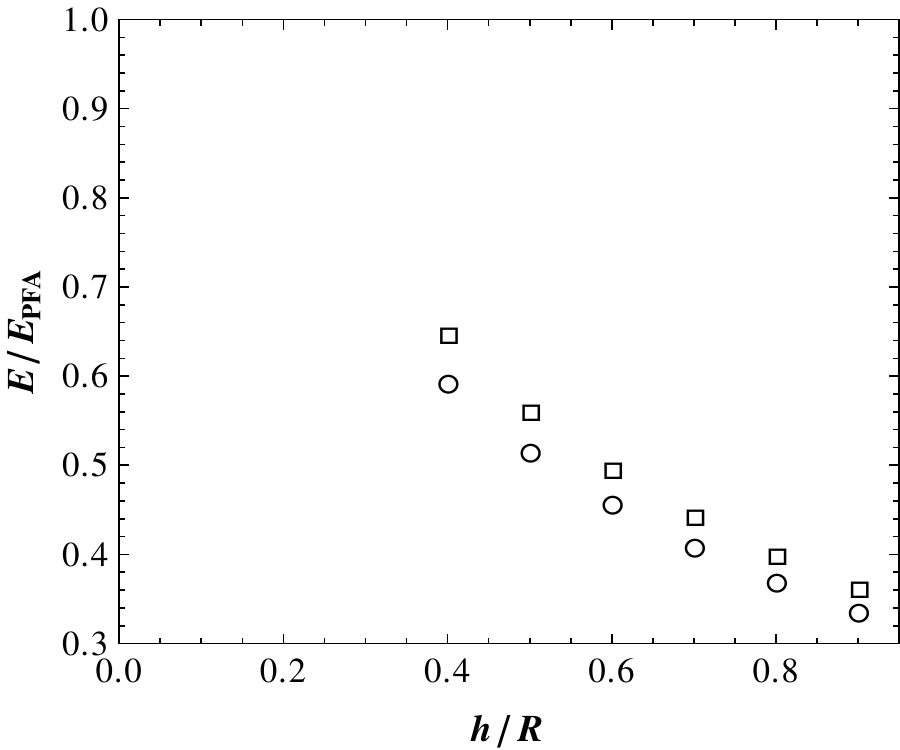} 
  \caption{Ratio of the numerical results for the Casimir energy shown
  in Fig.{~\ref{fig:numerics}} and the PFA energy 
  for perfect metal{s} (squares) and for plasma model with $\lambda_p/R=0.5$ (circles).
  The ratio is shown as a function of surface-to-surface distance $h$. Similar to Fig.{~\ref{fig:numerics}}, 
  the wire and particle have equal radiuii $R_\text{s}=R_\text{c}=R$.
}
  \label{fig:pfa}
\label{pfa-curve}
\end{figure}
For Drude and plasma objects, we use the Lifshitz {formula} \cite{Lifshitz56}, 
\begin{multline}
\label{eq:pfa-integral-plate}
\frac{{E}_{\rm plane}(h)}{A} = \frac{\hbar c}{(2\pi)^2} 
\int_0^\infty \kp^2 d\kp \int_1^\infty p\,dp \\
\times \ln\left[ (1-r^M_{\rm s}r^M_{\rm c}e^{-2\kp p h})(1-r^E_{\rm s}r^E_{\rm c}e^{-2\kp p h}) \right]\,, 
\end{multline}
with $p$ a dimensionless variable, $r_{\rm s}^{{M(E)}}$ and $r_{\rm c}^{{M(E)}}$ the Fresnel coefficients of {the 
surface elements for the} cylinder and sphere, respectively. The Fresnel coefficients of an object $a$ {are} given by
\bea
r^M_a(i c \kp,p) &=& \frac{\mu_a(i c\kp)-\sqrt{1+(n_a^2(i c\kp)-1)/p^2}}{\mu_a(i c\kp)+\sqrt{1+(n_a^2(i c\kp)-1)/p^2}}\,,\nonumber \\
r^E_a(i c \kp,p) &=& \frac{\epsilon_a(i c\kp)-\sqrt{1+(n_a^2(i c\kp)-1)/p^2}}{\epsilon_a(i c\kp)+\sqrt{1+(n_a^2(i c\kp)-1)/p^2}}\,,
\eea
where $n_a$ is the refraction index, $n_a(i c\kp)=\sqrt{\epsilon(i c \kp)\mu(i c \kp)}$.

The PFA energy is obtained using Eqs.~\eqref{eq:pfa-integral} and 
\eqref{eq:pfa-integral-plate} together with the dielectric
function {of Eq.~\eqref{di-fun}}.  Figure \ref{fig:pfa} 
shows the Casimir energy 
for a perfect metal ({open squares}) and plasma model 
$\lambda_{\rm p}/R=0.5$ ({open} circles) 
normalized to the PFA energy. 
The energies associated with the Drude model are not shown since they collapse 
on the data for the plasma model at short separations. 
{Our data show that as the distance between the sphere and the cylinder decreases, 
the values of the energies become closer to the PFA ones.}
Note that at short separations the energy converges with larger values of 
$\ell_{\rm max}$. 
Since the size of the matrix $\tilde{ \mathbb N}$ increases quadratically with $\ell_{\rm max}$
, the numerical calculation of the Casimir energy becomes 
extremely costly at short separations. In this work the interaction is calculated up to $h/R=0.4$. 

\section{Summary and Conclusions}

In summary, we have studied the Casimir energy 
between a cylindrical wire and a spherical particle. 
For large separations, we have derived the asymptotic energies for 
the Drude, plasma and perfect metal models. In addition, we have calculated 
the Casimir interaction between 
a metallic wire and an isotropic atom. 
Our results for the wire--atom system {is in} 
complete agreement with previous results obtained through a different method \cite{barash89,eberlein1,eberlein2,sc-mostep} . 

Furthermore, we have computed the Casimir interaction 
between a spherical particle and a wire. Such computations are quite demanding 
{due to lack of spherical symmetry. }
Our numerical results perfectly match the asymptotic energies.

For short separations, we obtained the energy using the 
Proximity Force Approximation (PFA) and compared it with our 
numerical data. This comparison indicates that as the distance 
between the wire and particle decreases, {the numerical results for the Casimir
energy becomes closer to the PFA one. }
{It is noteworthy that 
depending on the separation, the material properties of the 
metallic wire may not play {a} role in the interaction energy, 
similar to the parallel wires and wire--plate systems \cite{ours,ours1}. }

In a cylinder--sphere system, we do not observe a universal behavior 
as we have previously obtained for parallel wires and a wire--plate geometries because of the physical properties of the spherical particle.
However, { one} can still have ``universal" regimes in which the interaction does not 
depend on the material properties of the metallic wire. 

In case of the plasma wire with the plasma wavelength 
$\lambda_{\rm p}$ and radius $R_{\rm c}$, at sufficiently large separations, 
$d/R_{\rm c} \gg \exp(\lambda_{\rm p}^2/R_{\rm c}^2)$, 
the material properties of the wire 
does not play any role in the asymptotic interactions between the wire and a particle or an atom. \\
In contrast for the Drude wire with conductivity $\sigma$ 
and the characteristic length $\lambda_\sigma =2\pi c/\sigma$, at large separations 
$d^2/R_{\rm c}^2 \gg d/\lambda_{\sigma}$, 
the asymptotic energy depends on the 
the material properties of 
the wire. 
Quite interestingly, in the opposite limit, $d^2/R_{\rm c}^2 \ll d/\lambda_{\sigma}$ 
{and $d\gg \lambda_{\rm p}^2/\lambda_\sigma$}, 
the asymptotic interaction becomes  independent of the material properties of 
the Drude wire. The specific behavior of the Drude wires
{has been explained in Refs.~\cite{ours,ours1} in terms of large scale charge fluctuations.}
%

%
At the end we emphasize that since simple generic geometries appear 
in many nano- and micrometer--sized systems, the knowledge of the interaction between a metallic sphere and cylinder
could be {important} for an efficient 
design of low--dimensional structures. 

\acknowledgements
We thank M.~Kardar and U.~Mohideen for useful
discussions.  This work was supported by the
NSF through grants DMR-06-45668 (RZ), DARPA contract
No.~S-000354 (RZ and TE). \\
\appendix
\section{T-matrices}
\label{tmatrix}

In this appendix, for completeness, we present the T-matrices of a cylinder and a sphere, from Refs.~\cite{ours, roya10}.

The T-matrix of a sphere in spherical vector wave basis is diagonal in the quantum 
numbers $l$, $m$ and the electromagnetic polarizations $E$ and $M$ 
\begin{widetext}
\be
\label{eq:t-matrix-elem-sphere}
  T^{\textsc{m}}_{s,\ell m}= -\frac{\pi}{2} \frac{\eta I_{l+{1\over 2}}(\kp R)
\left[I_{l+{1\over 2}}(n\kappa R)+2n\kappa RI'_{l+{1\over 2}}(n\kappa R)\right] - n I_{l+{1\over 2}}(n\kappa R)
\left[I_{l+{1\over 2}}(\kappa R)+2\kappa R I'_{l+{1\over 2}}(\kappa R)\right]}
{\eta K_{l+{1\over 2}}(\kappa R)
\left[I_{l+{1\over 2}}(n\kappa R)+2n\kappa RI'_{l+{1\over 2}}(n\kappa R)\right] - n I_{l+{1\over 2}}(n\kappa R)
\left[K_{l+{1\over 2}}(\kappa R)+2\kappa R K'_{l+{1\over 2}}(\kappa R)\right]} \, ,
\ee
\end{widetext}
with $n=\sqrt{\epsilon(i\kappa)\mu(i\kappa)}$ and
$\eta=\sqrt{\epsilon(i\kappa)/\mu(i\kappa)}$.  The T-matrix elements
for E-multipoles, $T^{\textsc{e}}_{s,\ell m}$, are obtained from
Eq.~(\ref{eq:t-matrix-elem-sphere}) by interchanging $\epsilon$ and
$\mu$.

  The T-matrix elements 
of a cylinder with dielectric response $\epsilon(i\kp)$ and magnetic permeability $\mu(i\kp)$ are 
given by 
\begin{eqnarray}
   \label{eq:T-matrix-elements-ee}
  T^{EE}_{k_z n} &=& -\frac{I_n(pR)}{K_n(pR)} \frac{\Delta_2\Delta_3 +K^2}{\Delta_1\Delta_2+K^2}\,,\\
  \label{eq:T-matrix-elements-me}
  T^{EM}_{k_z n} &=& - \frac{K}{ \sqrt{\epsilon\mu} (pR)^2 K_n(pR)^2}
\frac{1}{\Delta_1\Delta_2 +K^2} \,,
\end{eqnarray}
 with $K = ({n k_z}/{(\sqrt{\epsilon\mu} R^2 \kappa)}) \left( {1}/{p'^2} -{1}/{p^2}\right)$
and 
\begin{eqnarray}
  \label{Delta}
  \Delta_1 &= & \frac{I'_n(p'R)}{p' R I_n(p'R)} -\frac{1}{\epsilon} \frac{K'_n(pR)}{pR K_n(pR)}\,.
\end{eqnarray}
Note that $\Delta_2$ can be obtained from Eq.~(\ref{Delta}) by interchanging 
$\epsilon$ with $\mu$, and $\Delta_3$ can be 
found by replacing $K'_n$  with $I'_n$ and $K_n$ with $I_n$ 
in Eq.~(\ref{Delta}). Moreover, $T^{MM}_{k_z n}$ can be obtained by replacing $\epsilon$ with $\mu$ 
and considering $T^{ME}_{k_z n}=-T^{EM}_{k_z n}$.

\section{T-matrix of a sphere in cylindrical basis }
\label{app:s-tmat-in-cyl}
The electromagnetic field far enough outside a sphere 
can be 
written in terms of the regular wave function with the electromagnetic 
polarization $P$,  $|E_{k_z m P}^{\rm reg}(ic\kappa)\rangle$, the free electromagnetic Green's function, ${\mathbb G}_0(ic\kappa)$, and the scattering operator of the sphere, ${\mathbb T}_{\rm s}(ic\kappa)$ \cite{universal}, 
\begin{equation}
\label{eq:farfield}
| E \rangle = |E_{k_z m P}^{\rm reg}\rangle - 
{\mathbb G}_0{\mathbb T}_{\rm s}|E_{k_z m P}^{\rm reg}\rangle\,,
\end{equation}
where $ic\kappa$ arguments are dropped for brevity. \\
The expansion of the  free Green's function in terms of the regular and outgoing wave functions is given by
\be
\label{eq:gf-expa}
{\mathbb G}_0 = \sum_{k_z m P} C_{P} |E_{k_z m P}^{\rm out}\rangle\langle E_{k_z m P}^{\rm reg}|\,,
\ee
where $C_E=-C_M={1 /{(2\pi L)}}$ is the normalization coefficient with $L$ 
the overall length of the cylinder. \\
Inserting Eq.~\eqref{eq:gf-expa} into Eq.~\eqref{eq:farfield} yields
\begin{equation}
\label{eq:farfield-tmat}
| E \rangle = |E_{k_z' m' P'}^{\rm reg}\rangle  +
\sum_{k_z m P} |E_{k_z m P}^{\rm out}\rangle \,T_{{\rm s},k_z m, k_z' m'} ^{PP'}\,,
\end{equation}
with
\be
\label{eq:Ts-c-basis}
T_{{\rm s},k_z m, k_z' m'} ^{PP'}= (-1) C_P \langle E^{\rm reg}_{k_z m P} |{\mathbb T}_{\rm s}|E^{\rm reg}_{k_z' m' P'}\rangle \,,
\ee
the T-matrix of a sphere in cylindrical basis.
Now we expand 
the cylindrical basis wave functions in terms of the spherical basis waves, 
\be
\label{eq:sol-exp}
| E^{\rm reg}_{k_z m P} \rangle = \sum_{\ell Q} D_{\ell mQ,k_z m P}| E^{\rm reg}_{\ell m Q}\rangle\,,
\ee
where $Q$ is the electromagnetic polarization, $\ell$ is the quantum number 
related to the spherical wave functions and the coefficients 
$D_{\ell mQ,k_z m P}$ are the elements of the conversion matrix 
from the cylindrical to spherical basis, see Appendix~\ref{conversion} for the 
detailed description. 
Since the azimuthal dependence of the wave functions in both sides of  Eq.~(\ref{eq:sol-exp}) 
are the same, the sum runs on the quantum number $\ell$ and polarization Q.\\
Inserting Eq.~(\ref{eq:sol-exp}) into Eq.~(\ref{eq:Ts-c-basis}), we obtain
\begin{multline}
\label{eq:Ts-c-basis2}
T_{{\rm s},k_z m, k_z' m'} ^{PP'}
= \sum_{\ell Q,\ell'Q'} \frac{C_P}{C'_Q(\kp)} D^{\dagger}_{k_z m P, \ell mQ} \\
\times T_{{\rm s},\ell m, \ell' m'} ^{QQ'} \,D_{\ell'm'Q',k_z' m' P'}\,,
\end{multline}
with 
\be
T_{{\rm s},\ell m, \ell' m'} ^{QQ'} = (-1)C'_Q(\kp)\langle E^{\rm reg}_{\ell mQ} | {\mathbb T}_{\rm s}|E^{\rm reg}_{\ell'm'Q'}\rangle\,,
\ee
the T-matrix of the sphere in the spherical basis, see Appendix \ref{tmatrix}, and 
$C'_M(\kp)=-C'_E(\kp)=\kp$ the normalization coefficients  
of the Green's function expansion in spherical basis. 
The ratio of the normalization coefficients in Eq.~(\ref{eq:Ts-c-basis2}) 
is $(1-2\delta_{P,Q})/(2\pi \kp L)$. Since the T-matrix of the sphere in spherical basis is 
diagonal in $l$, $m$ and polarization, Eq.~(\ref{eq:Ts-c-basis2}) is 
simplified to
\bea
\label{TsctoTss}
T_{{\rm s},k_z m,k_z'm}^{PP'} = 
\frac{1}{2\pi \kp L}\sum_{ \ell ={\rm max}(1,|m|)}^{\infty}\sum_{Q} (1-2\delta_{P,Q})& \nonumber\\
\times D^{\dagger}_{k_z m P, \ell mQ}  \,T_{{\rm s}, \ell m}^{Q}\,
D_{\ell mQ,k_z' m P'}\,.&
\eea

\section{Conversion Matrix ${\bf D}_{\ell m,k_z m}$}
\label{conversion}
The coefficients of the expansion of the cylindrical vector 
waves in terms of the spherical ones determine the elements of the conversion matrix 
${\bf D}_{\ell m,k_z m}$. These coefficients are known and have already been calculated \cite{samadar71, pog76,han08}. In this appendix we make the previously derived coefficients 
consistent with the Wick-rotated vector wave bases introduced in Ref.~\cite{universal}. 
The expansion of cylindrical vector waves $({\bf m}_{m\lambda},{\bf n}_{m\lambda})$ 
in terms of spherical vector waves  $({\bf m}_{\ell m},{\bf n}_{\ell m})$ is given by \cite{pog76}
\bea
\label{eq:conv-exp}
{\bf m}_{m\lambda} &=& \sum_{l=m}^{\infty} A_{\ell m\lam} {\bf m}_{\ell m} + B_{\ell m\lam}  {\bf n}_{\ell m}\,,\nonumber\\
{\bf n}_{m\lambda} &=& \sum_{l=m}^{\infty} A_{\ell m\lam} {\bf n}_{\ell m} + B_{\ell m\lam}  {\bf m}_{\ell m}\,,
\eea
where $\lam^2 = k^2-h^2$ and 
\bea
\label{eq:conv_old}
A_{\ell m\lam}&=&\frac{2\ell+1}{\ell(\ell+1)}\frac{(\ell-m)!}{(\ell+m)!} i^{\ell-m+1} k \sin(\al) 
\frac{d}{d\alpha} P_{\ell}^m(\cos(\al))\,, \nonumber\\
B_{\ell m\lam}&=&\frac{2\ell+1}{\ell(\ell+1)}\frac{(\ell-m)!}{(\ell+m)!} i^{\ell-m+1} m k
P_{l}^m(\cos(\al))\,,
\eea
with $\cos(\al)=h/k$. Note that $\bf m$ and $\bf n$ are the vector wave functions in Euclidean space. 

Taking into account $h \equiv k_z$, $k\equiv i \kp$ and $\lam \equiv i \sqrt{\kp^2+k_z^2}$,  the vector wave bases $({\bf M}^{\rm reg},{\bf N}^{\rm reg})$ defined in Ref.~\cite{universal} are related to the bases 
$({\bf m},{\bf n})$ defined in Ref.~\cite{pog76} by the relations
\bea
\label{eq:cyl-w-b}
{\bf m}_{m\lam}& =& i^m \sqrt{\kp^2+k_z^2}\, {\bf M}^{\rm reg}_{k_z m}(\kp,{\bf x})\,, \nonumber \\
{\bf n}_{m\lam} &=& i^{m-1} \sqrt{\kp^2+k_z^2}\, {\bf N}^{\rm reg}_{k_z m} (\kp,{\bf x})\,,
\eea
and 
\bea
\label{eq:sph-w-b}
{\bf m}_{\ell m}&=&i^\ell \sqrt{\frac{4\pi \ell(\ell+1)(\ell+m)!}{(2\ell+1)(\ell-m)!}} {\bf M}^{\rm reg}_{\ell m} (\kp,{\bf x})\,,\nonumber\\
{\bf n}_{\ell m}&=&i^{\ell-1} \sqrt{\frac{4\pi \ell(\ell+1)(\ell+m)!}{(2\ell+1)(\ell-m)!}} {\bf N}^{\rm reg}_{\ell m}(\kp,{\bf x})\,.
\eea
Plugging  Eq.~(\ref{eq:conv_old}) (after a Wick rotation ($k\rightarrow i\kp$)), 
Eq.~(\ref{eq:cyl-w-b}) and Eq.~(\ref{eq:sph-w-b}) into Eq.~(\ref{eq:conv-exp}), we obtain 
\bea
\begin{aligned}
{\bf M}_{k_z m}^{\rm reg}(\kp,{\bf x}) = 
\sum_{l=m}^{\infty} D_{\ell mM,k_z mM} {\bf M}_{\ell m}^{\rm reg} (\kp,{\bf x}) \\
 + D_{\ell mE,k_z mM} {\bf N}^{\rm reg}_{\ell m} (\kp,{\bf x})\,,\\
 {\bf N}_{k_z m}^{\rm reg}(\kp,{\bf x}) = 
 \sum_{l=m}^{\infty} D_{\ell mM,k_z mE} {\bf M}_{\ell m}^{\rm reg} (\kp,{\bf x}) \\ +
 D_{\ell mE,k_z mE} {\bf N}^{\rm reg}_{\ell m}(\kp,{\bf x})\,,
\end{aligned}
\eea
where the conversion matrix elements read

\begin{align}
\label{eq:conv}
D_{\ell mM,k_z mM} &=  (-1)^{\ell-m} \sqrt{\frac{4\pi(2\ell+1)(\ell-m)!}{\ell(\ell+1)(\ell+m)!}}
 \notag\\ 
&\times \left(1+\frac{k_z^2}{\kp^2}\right)^{1\over2} {{P'}_\ell^m} (-ik_z/\kp)\,, \notag\\
D_{\ell mE,k_z mM} &=  (-1)^{\ell-m} \sqrt{\frac{4\pi(2\ell+1)(\ell-m)!}{\ell(\ell+1)(\ell+m)!}}\,
 \\ 
&\times \,im \left(1+\frac{k_z^2}{\kp^2}\right)^{-{1\over2}}{P_\ell^m} (-ik_z/\kp)\,,\notag \\
D_{\ell mM,k_z mE} &= -D_{\ell mE,k_z mM}\,,\notag\\
D_{\ell mE,k_z mE} &= D_{\ell mM,k_z mM}\,. \notag
\end{align}

Since it is difficult to deal with the Legendre functions with complex 
arguments, we use the Rodrigues representation of Legendre polynomials and find
\be
\label{leg-real}
P_\ell^m(-i x) = (-i)^{\ell+m} f_\ell^m(x)\,,
\ee
where the real function $f_l^m$ is given by
\be
f_l^m(x) = \frac{1}{2^\ell \ell!} (1+x^2)^{m\over2} 
\frac{d^{\ell+m}}{dx^{\ell+m}} (1+x^2)^\ell \,.
\ee
Using Eq.~(\ref{leg-real}), we can write the conversion matrix $\mathbf D_{\ell m, k_z m}$ 
in terms of a modified matrix $\mathbf {\tilde D}_{\ell m, k_z m}$,
\be
{\bf D}_{\ell m,k_z m} = (-1)^{\ell-m} (-i)^{\ell+m-1} {\bf \tilde D}_{\ell m,k_z m}\,,
\ee
with
\begin{align}
{\tilde D}_{\ell m M,k_z m M}& = \sqrt{\frac{4\pi(2\ell+1)(\ell-m)!}{\ell(\ell+1)(\ell+m)!}} \notag \\
&\times \left(1+\frac{k_z^2}{\kp^2}\right)^{1\over2}{f_\ell^m}' (k_z/\kp)\,, \notag\\
{\tilde D}_{\ell m E,k_z m M} &= \sqrt{\frac{4\pi(2\ell+1)(\ell-m)!}{\ell(\ell+1)(\ell+m)!}}\\
&\times m\left(1+\frac{k_z^2}{\kp^2}\right)^{-{1\over2}}{f_\ell^m}(k_z/\kp)\,, \notag \\
{\tilde D}_{\ell mM,k_z mE} & =  -{\tilde D}_{\ell mE,k_z mM}\,,\notag\\
{\tilde D}_{\ell mE,k_z mE} & =  {\tilde D}_{\ell mM,k_z mM}\,.\notag
\end{align}

\bibliography{ref.bib}

\end{document}